\documentclass[usenatbib]{mn2e}
\usepackage[dvips]{color}
\usepackage{ulem}
\usepackage{longtable}
\usepackage{url}

\usepackage{graphicx}
\usepackage{lscape}
\def\arcdeg{\hbox{$^\circ$}}
\def\arcmin{\hbox{$^\prime$}}

\def\flux{erg s$^{-1}$ cm$^{-2}$}
\def\phflux{ph\,cm$^{-2}$\,s$^{-1}$}

\def\apj{ApJ}

\def\aap{A\&A}

\def\nat{Nature}

\newcommand {\be}{\begin {equation}}
\newcommand {\ee}{\end {equation}}

\def\fl{ph\,cm$^{-2}$\,s$^{-1}$}
\title[Galactic survey of $^{44}$Ti sources with {\it INTEGRAL}]
{Galactic survey of $^{44}$Ti sources with the IBIS telescope onboard {\it INTEGRAL}}
\author[Tsygankov et al.]{Sergey S.\,Tsygankov,$^{1,2}$\thanks{E-mail:stsygankov@gmail.com}
Roman A.\,Krivonos,$^{2}$ 
Alexander A.\,Lutovinov,$^{2,3}$ 
\newauthor
Mikhail G.\,Revnivtsev,$^{2}$
Eugene M.\,Churazov,$^{4,2}$
Rashid A.\,Sunyaev$^{4,2}$
\newauthor
and Sergey A. Grebenev$^{2}$\\
$^1$ Tuorla Observatory, Department of Physics and Astronomy, University of Turku, 
  V\"ais\"al\"antie 20, FI-21500 Piikki\"o, Finland \\
$^{2}$ Space Research Institute of the Russian Academy of Sciences, Profsoyuznaya Str. 84/32, Moscow
  117997, Russia\\
$^{3}$ Moscow Institute of Physics and Technology, Moscow region, Dolgoprudnyi, Russia \\
$^{4}$ Max Planck Institute for Astrophysics, Karl-Schwarzschild Str. 1, Garching, 85741, Germany
}

\begin{document}

\date{Accepted 2016 March 3. Received 2016 March 3; in original form 2015 December 11}

\pagerange{\pageref{firstpage}--\pageref{lastpage}} \pubyear{2015}

\maketitle

\label{firstpage}

\begin{abstract}
{We report the results of the deepest Galactic Plane ($|b| <
  17.5^{\circ}$) survey in the 67.9 and 78.4 keV nuclear de-excitation
  lines of titanium-44 ($^{44}$Ti) performed using the data acquired
  with the IBIS/ISGRI instrument onboard the \textit{INTEGRAL}
  satellite during 12 years of operation.  The peak sensitivity of our
  survey reached an unprecedented level of $4.8\times10^{-6}$ \fl\
  ($3\sigma$) that improves the sensitivity of the survey done by {\it
    CGRO}/COMPTEL by a factor of $\sim5$. As a result, constraining
  upper limits for all sources from the catalog of Galactic supernova
  remnants \citep[SNRs; ][]{green2014} are derived. These upper limits
  can be used to estimate the exposure needed to detect $^{44}$Ti
  emission from any known SNR using existing and prospective X- and
  gamma-ray telescopes. Among the youngest Galactic SNRs, only Cas A
  shows significant $^{44}$Ti emission flux in good agreement with the
  {\it NuSTAR} measurements. We did not detect any other sources of
  titanium emission in the Galactic Plane at significance level higher
  than 5$\sigma$ confirming previous claims of the rarity of such
  $^{44}$Ti-producing SNRs.  }

\end{abstract}

\begin{keywords}
gamma-rays: ISM; supernova remnants; nuclear reactions, nucleosynthesis, abundances
\end{keywords}

\section{Introduction}

Supernova (SN) explosions determine the chemical and physical
evolution of the Universe. An understanding of their mechanism is the
keystone of many branches of the modern astrophysics. However, there
is no way to observe directly physical processes during the first
stages of the explosion due to a very high opacity. The most promising
solution of this problem is to measure the amount of radioactive
elements synthesised during SN explosion and based on that to verify
the existing models \citep[see, e.g.,][]{2012A&ARv..20...49V}.

Probably the best candidate to such explosion tracer is radioactive
isotope of titanium-44 ($^{44}$Ti). It has a lifetime of about 85 years
\citep[e.g.,][]{2006PhRvC..74f5803A}, that is long enough to guarantee
the substantial fraction of the synthesised titanium to remain active
after the envelope become transparent. Such lifetime is much shorter
than for, e.g., $^{26}$Al ($\tau\sim10^{6}$ yr), that permits to
associate the measured flux with a specific supernova remnant (SNR),
thus increasing chances to detect young (few hundred years) SNR.

$^{44}$Ti is produced during SN explosions deep inside the progenitor
star and its yield is very sensitive to the particular conditions
there. Mainly it is synthesised during core collapse SN within an
$\alpha$-rich freeze-out, and, hence, depends strongly on the
expansion speed of the inner layers of the ejecta
\citep{1996snih.book.....A,2010ApJS..191...66M}. Other factors affecting yield of
$^{44}$Ti are the mass cut between the proto-neutron star and the
ejecta, progenitor mass and metallicity, explosion asymmetry
\citep{1998ApJ...492L..45N}.  Core collapse SNe produce typically
$10^{-5}$--$10^{-4}$ M$_{\odot}$ of titanium
\citep[e.g.][]{1996ApJ...464..332T}. The expected yield of $^{44}$Ti
in type Ia SNe is ranging from $\sim10^{-6}$ M$_{\odot}$ for a
centrally ignited pure-deflagration to $\sim6\times10^{-5}$
M$_{\odot}$ for an off-center delayed detonation
\citep{1999ApJS..125..439I,2010ApJ...712..624M}. However, in the
so-called double-detonation sub-Chandrasekhar model it can reach
$\sim10^{-3}$ M$_{\odot}$ \citep[][]{2010A&A...514A..53F}.

From the observational point of view the signatures of titanium decay
through the chain $^{44}$Ti$\rightarrow^{44}$Sc$\rightarrow^{44}$Ca
can be observed in broad range of energies -- it produces on average
$\sim$0.17, $\sim$0.88, $\sim$0.95 and $\sim$1 photons per one decay
in the lines at energies 4.1, 67.9, 78.4, and 1157 keV, respectively.
Hence, it could be detected by telescopes based on completely
different principals (an overview of previous results is presented in
Section \ref{hist}).

Based on the comparison of the survey done by COMPTEL telescope
onboard {\it Compton Gamma-Ray Observatory (CGRO)} in the 1.157 MeV
line \citep{1997A&A...324..683D,1999ApL&C..38..383I} with theoretical
expectations \cite{2006A&A...450.1037T} concluded that SNe producing
$^{44}$Ti are not typical events. Particularly, they expect 5-6
positive detections of $^{44}$Ti sources above COMPTEL's sensitivity
limit ($\sim10^{-5}$ \fl\ at $1\sigma$ level). However at that moment
Cas A was the only confidently detected source of $^{44}$Ti emission
\citep{1994A&A...284L...1I}. It is worth noting that more recent
calculations performed by \cite{2013ApJ...775...52D} give results
consistent with one detected $^{44}$Ti source in the survey with
sensitivity reached by COMPTEL. Only recently, thanks to significant
exposures collected by hard X-ray space telescopes currently operating
in orbit (\textit{INTEGRAL}/IBIS, \textit{Swift}/BAT,
\textit{NuSTAR}), new evidence of SNRs emitting in $^{44}$Ti lines
started to appear, in particular SN 1987A
\citep{2012Natur.490..373G,2015Sci...348..670B} and Tycho's SNR
\citep{2014ApJ...797L...6T}.

One of the main goal of the current paper is to search for new and
previously unknown $^{44}$Ti emission sources, serendipitous
detections of which could provide us information about SN activity in
the Milky Way within the last few centuries. Last systematic survey
for $^{44}$Ti sources in MeV domain was conducted by COMPTEL in the
work of \cite{1997A&A...324..683D} and \cite{1999ApL&C..38..383I}. The
authors did not detect new unknown $^{44}$Ti sources that was in
agreement with canonical values of 2.5 to 3 Galactic SN events per
century. Later, \cite{2004ESASP.552...81R} confirmed the absence of
new sources of $^{44}$Ti line emission in the first survey of the
Galactic Centre region ($|l| < 30^{\circ}$) using the {\it INTEGRAL}
IBIS/ISGRI data from the first year of operation.

Core-collapse SNe can be embedded in the dense molecular clouds that
gave birth to their massive progenitors. The high column densities of
the Galactic, local molecular cloud, and even circumstellar
obscuration may prevent detection of such a recent SN
events. Undetectable at optical wavelength, young SNRs could be
revealed through the decay of $^{44}$Ti in emission lines at 67.9 and
78.4 keV thanks to high penetrating power of hard X-rays. As shown by
recent detection of $^{44}$Ti from SN~1987A in the Large Magellanic
Cloud (LMC) by \cite{2012Natur.490..373G}, {\it INTEGRAL} IBIS/ISGRI
has the potential for detecting these sources even in the nearby
galaxies.

With sensitivity achieved in the current 12 years survey by {\it
  INTEGRAL}/IBIS in the Galactic Plane (factor of 2--5 better than
COMPTEL) we can improve our knowledge about recent SN activity in the
Galaxy by model independent and systematic-free imaging at better
angular resolution (12\arcmin) in comparison to the COMPTEL experiment
($\sim$1$^{\circ}$), which is, however still not enough to resolve
spatial morphology of SNRs with typical size of a few arcmins.

In this paper we utilize main advantages of the \textit{INTEGRAL}/IBIS
telescope (large field of view, high sensitivity and energy
resolution, huge amount of collected data) to systematically search
for $^{44}$Ti emission (in the two low-energy lines at 67.9 and 78.4
keV) from the Galactic SNRs, selected at $|b| < 17.5^{\circ}$.

\section{Observations and Data Analysis}

The {\it INTEGRAL} observatory \citep{win03} has demonstrated a great
success in surveying the sky in hard X-rays at energies above
20~keV. Large field of view of $28^{\circ}\times28^{\circ}$, moderate
angular resolution of a few arcmins, and one of the highest
sensitivity for the class of its optical design (coding aperture) led
to many relevant survey papers (see, e.g., \citealt{krietal2012} and
references therein).  Thanks to effective work of the IBIS coded-mask
telescope \citep{ibis} over more than ten years, we can conduct the
first imaging survey of the whole Galactic Plane in titanium emission
lines, significantly extending the work by \cite{2004ESASP.552...81R}
in the Galactic Centre. The fact that {\it INTEGRAL} spent most of its
observational time towards the Galactic Plane, where most of the SNRs
reside, makes the current survey unprecedented in sensitivity and
coverage.

For the current analysis we utilized all the publicly available data
taken until October, 2014 (or spacecraft revolution 1469). The data
were screened and reduced in accordance with our previous surveys (see
e.g. \citealt{2005MNRAS.357.1377C,churazov2014,kri2010a} and
references therein). To take into account the long-term ISGRI detector
degradation and subsequent decrease in the efficiency
\citep{caballero} we corrected for a secular gain variations and
adjusted the efficiency in each IBIS/ISGRI energy bands using the flux
of the Crab nebula measured in the observation that is closest in time
\citep[for details see][]{krietal2012}. This method corresponds to a
smooth recalibration of the ancillary response function (ARF) over the
time span of the survey.  In our IBIS/ISGRI data analysis we use the
diagonal energy redistribution matrix designed to reproduce Crab
spectrum in the form 
\be\label{eq1} 
{\rm d}N/{\rm d}E=10.0\times E_{\rm keV}^{-2.1}~{\rm phot}~{\rm cm}^{-2}~{\rm s}^{-1}~{\rm
  keV}^{-1} 
\ee 
which is a good representation of the historic Crab
observations \citep[see, e.g.,][]{2007A&A...467..529C}.

The total list of the available data over the full sky contains 99108
individual {\it INTEGRAL} pointings with typical exposure of 2~ks, or
so called \textit{Science Window}s (\textit{ScW}s), which corresponds
to $\sim178$~Ms of the effective exposure. The data of the ISGRI
detector layer \citep{isgri} for each ScW were converted into sky
images in the energy bands of interest (defined below). Similar to the
{\it INTEGRAL} nine-year Galactic Plane survey by \cite{krietal2012},
the survey mosaicing was organized in six overlapping
$70^{\circ}\times35^{\circ}$ Galactic cartesian map projections
centered at $b=0^{\circ}$ and $l=0^{\circ}$ (GC), $\pm50^{\circ}$,
$\pm115^{\circ}$, and $l=180^{\circ}$ (Galactic anticenter). The
latitude coverage of the current survey $|b| < 17.5^{\circ}$ was
chosen to take advantage of the large IBIS field of view and {\it
  INTEGRAL} observational pattern in the Galactic Plane. The final
data set used in this work comprises 62509 \textit{ScW}s or 104~Ms of
dead-time corrected exposure, which corresponds to about 60\% of the
full data set.

The search for the $^{44}$Ti lines at 67.9 and 78.4 keV imposes severe
requirements on energy calibration precision of the instruments. We
used the most up-to-date ISGRI energy reconstruction available through
the Offline Scientific Analysis (OSA) version 10.1, provided by
ISDC\footnote{ISDC Data Centre for Astrophysics,
  http://www.isdc.unige.ch/}. As described by \cite{caballero}, the
energy resolution of ISGRI at tungsten (W) fluorescent line located at
58.8297~keV shows gradual broadening from $\sim5$ to $\sim10$~keV
(FWHM) over 10 years of the current mission life span, which still
makes it possible to study narrow $^{44}$Ti emission lines. To trace
continuum and line emission we defined working energy bands as shown
in Table~\ref{tab:ebands}. Two $^{44}$Ti lines at 67.9 and 78.4 keV
are accommodated in the energy bands E3 and E5, respectively, and both
lines are placed in E7 band. E5 is wider than E3 by 1~keV owing to an
initial ISGRI energy resolution of 8\% at 60~keV \citep{isgri}.
According to our measurements, the average energy resolution (FWHM) at
60 keV over the survey data span is $\sim7.5$~keV, which is comparable
to the width of the selected energy bands and consistent with
\cite{caballero}. To prevent any losses in the $^{44}$Ti line emission
due to the finite energy resolution as a reference energy band we took
much broader E7 band containing both lines.

%=================================================================
\begin{table}
\noindent
\centering
\caption{Working energy bands for the current $^{44}$Ti survey.}\label{tab:ebands}
\centering
\vspace{1mm}
  \begin{tabular}{|c|c|r|c|c|c|c|}
\hline\hline
Name & Range & Width & 1~mCrab & $^{44}$Ti \\
     & [ keV ] & [ keV ] & [ \flux ] & lines \\
     &         &         & [ \phflux ] &    \\
\hline
E1 & $25.0-40.0$& 15.0 & $5.35\times10^{-12}$ & \\
   &            &      & $1.07\times10^{-4}$ & \\
E2 & $40.0-64.6$ & 24.6 & $5.19 \times10^{-12}$ & \\
   &             &      & $6.45 \times10^{-5}$ & \\
E3 & $64.6-71.2$ & 6.6 & $1.02 \times10^{-12}$ & 67.9~keV\\ 
   &             &     & $9.48 \times10^{-6}$ & 67.9~keV\\ 
E4 & $71.2-74.6$ & 3.4 & $0.48 \times10^{-12}$ & \\
   &             &     & $4.14 \times10^{-6}$ & \\
E5 & $74.6-82.2$ & 7.6 & $1.00 \times10^{-12}$ & 78.4~keV \\
   &             &     & $8.01 \times10^{-6}$ & 78.4~keV \\
E6 & $82.2-130.0$ & 47.8 & $4.65 \times10^{-12}$ & \\
   &              &      & $2.84 \times10^{-5}$ & \\
E7 & $64.6-82.2$ & 17.6 & $2.52 \times10^{-12}$ & both \\ 
   &             &      & $2.17 \times10^{-5}$ & both \\ 
\hline
\end{tabular}
\vspace{3mm}
\end{table}
%=================================================================

Despite the great improvement of the ISGRI detector absolute energy
reconstruction in OSA 10.1 compared to previous versions, the
$\sim1.5\%$ gain variations during each orbit are still present, which
lead to $\sim1$~keV uncertainty at $\sim80$~keV. We applied additional
gain correction based on the position of the tungsten
fluorescent line, suppressing the uncertainty of energy reconstruction
for the detector-averaged line centroid to less than 0.1~keV.

Table~\ref{tab:ebands} contains flux conversion coefficients between
different flux units (mCrab and \flux, and \phflux) in our working
energy bands assuming Crab spectrum in the form given by
eq. (\ref{eq1}).

\section{Results}

\subsection{The search for new sources of  $^{44}$Ti line emission}
\label{section:search}

%=================================================================
\begin{figure}
\includegraphics[width=\columnwidth]{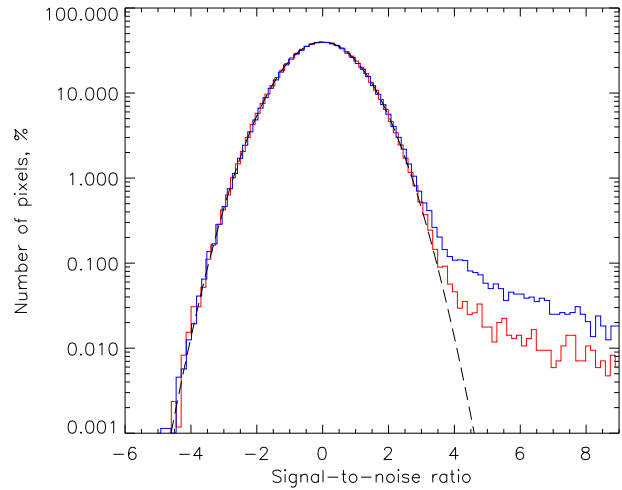}

\caption{Distribution of signal-to-noise ratio values of a number of
  pixels in $70^{\circ}\times 35^{\circ}$ sky map centered at
  $l=0^{\circ}$ (blue line) and $l=+50^{\circ}$ (red line) in E7
  energy band (see Table~\ref{tab:ebands}). The long-dashed line
  represents normal distribution with zero mean and unit
  variance.}\label{fig:snr_distr}

\end{figure}
%=================================================================

The search for new sources has been performed on each projected mosaic
images in the reference energy band E7 ($64.6-82.2$~keV) containing
both $^{44}$Ti lines and therefore providing better statistics. We
defined detection threshold allowing not more than one false detection
assuming pure photon counting statistics. Given the angular resolution
of the IBIS telescope (12\arcmin), the Galactic survey ($|b| <
17.5^{\circ}$) contains $\sim3\times10^5$ independent pixels which can
generate one statistical fluctuation at $4.7\sigma$. We should stress
that the sensitivity of our $^{44}$Ti survey is limited by count
statistics only and not affected by systematics, which seriously
limits the sensitivity of IBIS telescope at lower energies, especially
in the crowded region of the Galactic Centre \citep[see,
  e.g.,][]{kri2010b,krietal2012}. Fig.~\ref{fig:snr_distr} shows a
signal-to-noise distribution of pixel values in the sky mosaic at
$l=0^{\circ}$ (blue line) and $l=+50^{\circ}$ (red line) along with
the normal distribution representing a statistical noise.  Note that
the positive tail of this distribution is formed by continuum
sources. As seen from the figure (overall shape and, in particular,
negative side of the distribution), the observed noise distribution
can be very well described by normal distribution with zero mean and
unit variance.

%=================================================================
\begin{figure*}
 \includegraphics[width=0.98\textwidth]{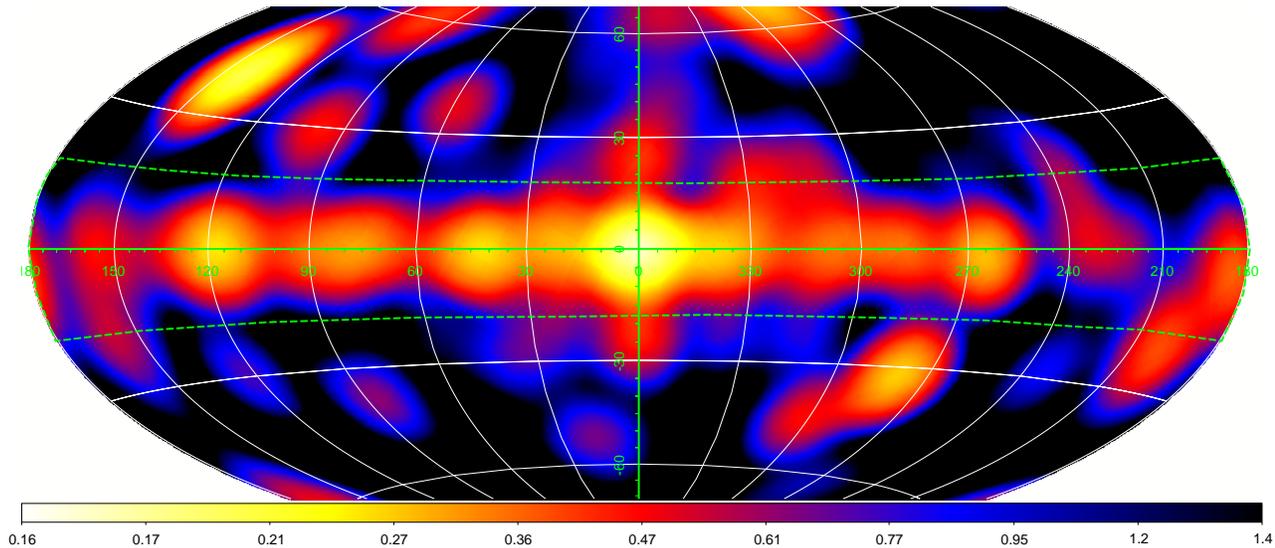}
 \caption{The map shows $1\sigma$ flux sensitivity in 64.6--82.2 keV
   energy band (E7; Table~\ref{tab:ebands}) ranging from 0.16 to 1.4
   mCrab. Two dashed green lines comprise geometrical area of the survey
   ($|b|<17.5^{\circ}$).}\label{fig:sen}
\end{figure*}
%=================================================================

\begin{figure}
 \includegraphics[width=0.48\textwidth]{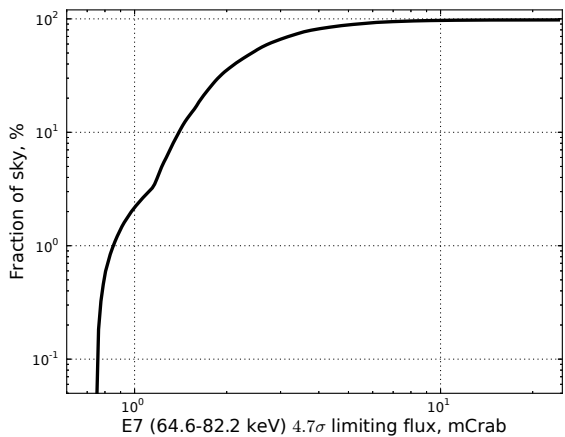}
 \caption{Fraction of the sky covered as a function of $4.7\sigma$
   limited flux. 100\% corresponds to geometrical area of the survey
   (12680~deg$^{2}$).}\label{fig:area}
\end{figure}

The source detection sensitivity of our $^{44}$Ti survey is not
completely uniform over the Galactic Plane (Fig. \ref{fig:sen}). The
maximum IBIS/ISGRI sensitivity is achieved in the region of the
Galactic Center (GC) which exhibit the largest exposure accumulated
over 12 years. The survey sky coverage in 64.6--82.2 keV energy band
(E7) as a function of a $4.7\sigma$ limiting flux is demonstrated in
Fig.~\ref{fig:area}. The peak sensitivity is about 0.7~mCrab
($1.8\times 10^{-11}$ \flux\ in E7 band) in the GC.  Regarding the
range of the Galactic latitudes $|b| < 17.5^{\circ}$, the survey
covers 10\% of its geometrical area (12680 deg$^{2}$) at 1.4~mCrab
($3.5\times 10^{-11}$ \flux) and 90\% at 5.4~mCrab ($1.4\times
10^{-10}$ \flux).

%=================================================================
%=================================================================

\begin{figure}
\includegraphics[width=\columnwidth,bb=37 180 545 575, clip]{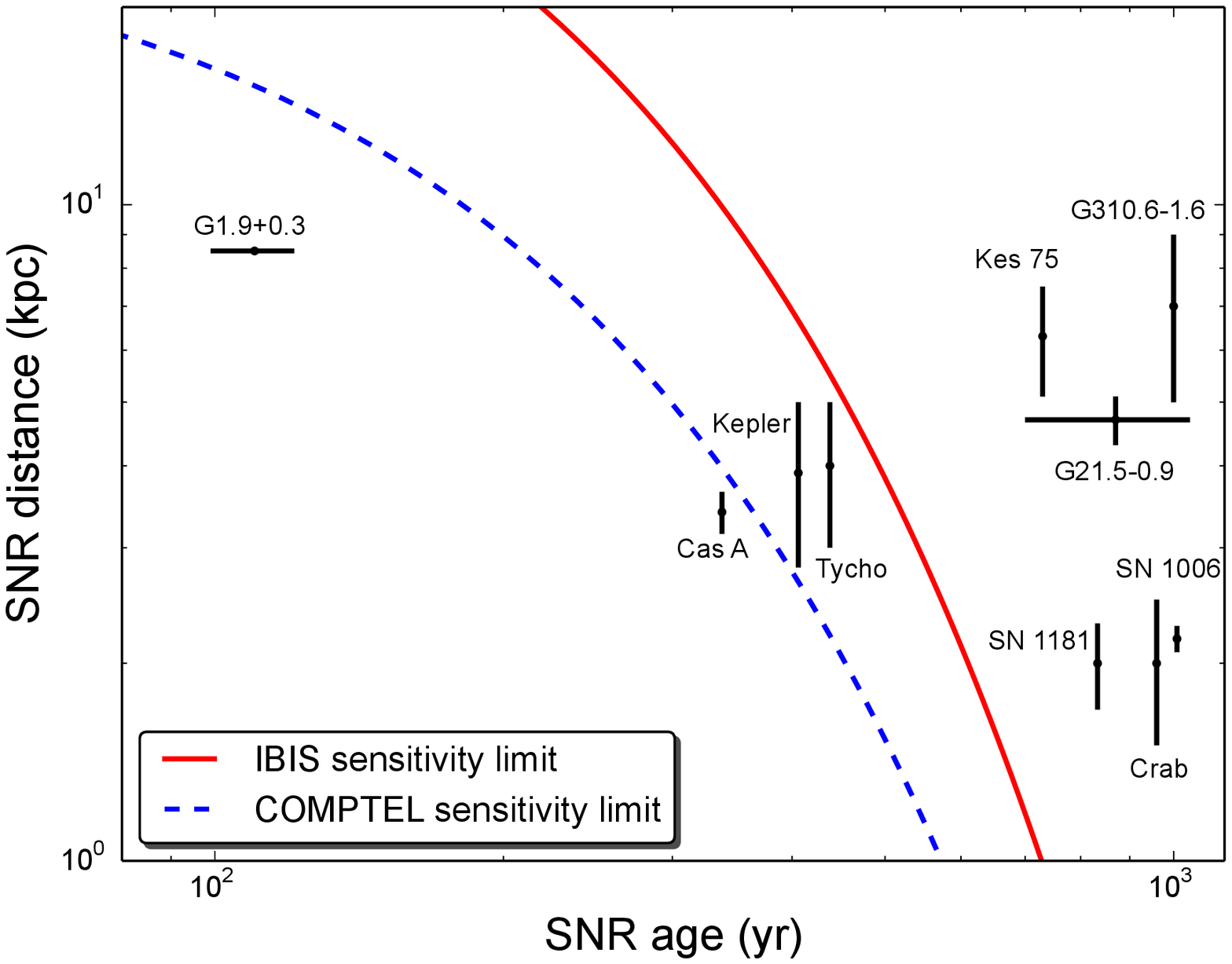}

\caption{Age-distance diagram showing the capability of $^{44}$Ti
  emission detection using the COMPTEL and current Galactic
  \textit{INTEGRAL}/IBIS surveys. The red solid line and blue dashed
  line show $1\sigma$ peak sensitivity reached in the current
  \textit{INTEGRAL}/IBIS survey ($\sim1.7\times10^{-6}$\fl) and
  COMPTEL survey ($9\times10^{-6}$\fl), correspondingly.  A $^{44}$Ti
  lifetime of 85.0 yr and a yield of $1\times10^{-4}$ M$_{\odot}$ were
  assumed.}\label{fig:diag}

\end{figure}

The peak sensitivity of our survey reached in the region of the
Galactic centre is $1.0\times10^{-5}$ \phflux\ ($3\sigma$) in E7 energy
band, containing both 67.9 keV and 78.4 keV lines. Taking into account
the approximate equality of intensities in these two lines the
$3\sigma$ sensitivity per one $^{44}$Ti line in the current {\it
  INTEGRAL}/IBIS survey can be recalculated as $\sim4.8\times 10^{-6}$
\phflux. According to the different efficiencies of emission in
different lines the corresponding sensitivity in the 1.157 MeV line is
$\sim5.2\times 10^{-6}$ \phflux.  This is factor of $\sim5$ better
than sensitivity of $2.7\times10^{-5}$ \phflux\ in the 1.157 MeV line
($3\sigma$) reached with the COMPTEL experiment in the Galactic Plane
\citep{1997A&A...324..683D,1999ApL&C..38..383I}. Regarding the
Galactic Plane at $|l| < 50^{\circ}$ a factor of $\sim2$ improvement
in sensitivity is achieved. To illustrate the improvement of
sensitivity in the current survey in comparison to the COMPTEL one, we
constructed the age-distance diagram assuming a $^{44}$Ti yield of
$Y_{44}=1\times10^{-4}$ M$_{\odot}$ (see Fig. \ref{fig:diag}; ages and
distances for the plotted SNRs are listed in Table \ref{tab:hist}).

The list of candidate sources detected with $S/N>4.7\sigma$ and within
latitude range of the current survey ($|b| < 17.5^{\circ}$) contains
only one previously unknown object, the source IGR~J$18406+0451$ found
at position R.A.=18h~40m~40.80s, Decl.=04d~51m~36.0s (equinox J2000),
which is 15\arcmin\ away from the known low mass X-ray binary Serpens~X-1
\citep[][listed as 3A 1837+049]{liu2007}. The source was registered
with a flux of $1.4\pm0.28$~mCrab ($5.0\sigma$ detection) in the E7
energy band. The localization accuracy of the sources detected with
IBIS/ISGRI depends on the source significance
\citep{gros2003,2006ApJ...636..765B}. According to \cite{krietal07},
the positional $1\sigma$ uncertainty of the source detected at
$S/N\simeq5-6\sigma$ is 2.1\arcmin.

To further investigate the properties of IGR~J18406+0451 we
reconstructed its flux in all energy bands E1-E7.  IGR~J18406+0451
does not show significant detection in E1 ($<0.23$~mCrab, $3\sigma$);
in E2 ($<0.35$~mCrab, $3\sigma$); and in E6 ($<0.84$~mCrab,
$3\sigma$) but exhibits marginal
detection in E3 ($1.09\pm0.39$~mCrab, $2.8\sigma$); E4
($1.51\pm0.69$~mCrab, $2.19\sigma$), and E5 ($1.63\pm0.51$~mCrab,
$3.2\sigma$).

Assuming IGR~J18406+0451 is a $^{44}$Ti emission source, the combined
probability of $^{44}$Ti line detection in E3 and E5 is
$7\times10^{-6}$ ($\sim4.5\sigma$), which is lower than the E7
significance ($5.0\sigma$). This tells us that intermediate energy
band E4 can contain some contribution from $^{44}$Ti lines probably
due to the ISGRI spectral resolution (5--10 keV) at these
energies. This assumption is confirmed by the combined probability of
signal detection in E3/E4/E5 bands, which is $7\times10^{-6}$ or
$5.2\sigma$.

In all energy bands of the continuum emission E1, E2 and E6,
IGR~J18406+0451 is below the detection threshold, while the source
shows significant ($\sim5\sigma$) emission in E7 energy band
containing two $^{44}$Ti emission lines, and this excess is dominated
by the flux in E3 and E5 bands, where the two low-energy $^{44}$Ti
lines are located.  Owing to a relatively high Galactic latitude of
4.6$^{\circ}$, we can expect the source candidate IGR~J18406+0451 to
be nearby, in particular if it is the remnant of a core-collapse SN,
potentially located in the Sagittarius spiral arm at 2--3 kpc
\citep{2009ApJ...693..419Z}.

In spite of a relatively high S/N ratio for this new source it is still not
far from the detection threshold ($4.7\sigma$) corresponding to the
registration of one false source in our survey. Therefore additional
observational evidence are required to establish if the source is real
or not.

For this purpose, we asked for DDT observations of this region with
the {\it XMM-Newton} observatory. These observations were performed on
October 9, 2015 with a total exposure of $\sim30$ ksec. Data were
reduced with the standard SAS
software\footnote{\url{http://xmm2.esac.esa.int/sas/}}. No bright
sources were detected inside the {\it INTEGRAL} error circle. The
search for faint sources met some difficulties due to the illumination
from the bright nearby source Serpens X-1. A conservative point-source
upper limit ($3\sigma$) on the flux at the IGR~J18406+0451 position is
about $8\times10^{-14}$ \fl\ in the 0.5-10 keV energy band (under the
assumption of a power-law spectrum with slope $\Gamma=-2$).

It is important to note, that apart from hard X-ray emission lines at
67.9 and 78.4 keV, there is another one originated from the $^{44}$Ti
decay -- the $^{44}$Sc fluorescence line with the $^{44}$Ti emission
in this line (with a branching ratio of 0.17 per decay). The averaged
nuclear de-excitation lines flux measured from IGR\,J18406+0451
$F_{44}=1.27\times10^{-5}$ \phflux\ can be translated into the
expected flux in the 4.1 keV line $F_{4.1 keV}=2.16\times10^{-6}$
\phflux. An upper limit ($3\sigma$) for the flux in the narrow energy
band 3.9-4.3 keV, derived from the {\it XMM-Newton} data, is
$\simeq8\times10^{-7}$ \phflux, that is several times lower than the
above estimations.  Note that the expected flux in the 4.1 keV line is
true for the fully transparent SNR envelope and with no additional
absorption by the ISM gas along the line of sight at these soft X-ray
energies.

The obtained results can be interpreted in two ways: the source
candidate IGR\,J18406+0451 is a false detection or its emission in
soft X-rays is strongly absorbed. Thus, additional observations in
hard X-rays by instruments with a better sensitivity and angular
resolution (like {\it NuSTAR}) are required to confirm the detection
of IGR\,J18406+0451.

\subsection{A catalogue of Galactic supernova remnants}
\label{green}

Thanks to the wide field of view and good angular resolution of the
IBIS telescope we can for the first time put upper limits on the
$^{44}$Ti line emission from all known Galactic SNRs. For this
purpose we used the catalog of Galactic SNRs presented in
\cite{green2014}. The catalogue contains 294 sources revealed through
radio and infra-red surveys, as well as X-ray observations. Most of
the SNRs are old, and generally we do not expect to detect significant
signal from them, but no systematic search for $^{44}$Ti emission has
been performed so far. Since we are looking at known sky positions, we
can go below detection threshold of the current survey ($4.7\sigma$)
and provide marginal detection or conservative upper limit.

Table~\ref{tab:green} lists measured fluxes or $3\sigma$ upper limits
towards 294 SNRs from \cite{green2014}. As indicated in the notes to
the table, 60 SNRs are subject to spatial confusion within
12\arcmin\ (the IBIS/ISGRI angular resolution) from known hard X-ray
sources. The measured flux of 9 SNRs is most probably dominated by the
continuum emission of the associated X-ray source, rather than by
putative $^{44}$Ti line emission.  If significance of the detection of
such source is above $3\sigma$ level, an upper limit ($3\sigma$) in
the 78.4 keV line is also shown assuming a simple power-law continuum
model. Marginal detection (at less than 3$\sigma$) of the $^{44}$Ti
line emission and/or continuum is registered from Tycho SNR (see
Sec.\ref{hist}), G065.7+01.2, G069.0+02.7, G315.1+02.7 and
G356.3--01.5.

%=================================================================
\begin{table*}
\centering
\caption[A sample of the catalogue of Galactic SNRs \citep{green2014} with measured fluxes or
$3\sigma$ upper limits in the $64.6-82.2$~keV band] {A sample of the catalogue of Galactic SNRs \citep{green2014} with measured fluxes or
$3\sigma$ upper limits in the $64.6-82.2$~keV band. The full version is available in the online version of this article.} \label{tab:green}
 \begin{tabular}{@{}clrrccccl}
  \hline
\hline
No. & Name\footnotemark[1] & l$^{II}$ & b$^{II}$ & Flux$_{\rm 64.6-82.2~keV}$,\footnotemark[2] &  Notes\footnotemark[3] \\
    &                      & deg     & deg     & $10^{-12}$ erg cm$^{-2}$ s$^{-1}$ &   \\ \hline
001 & SNR G000.0+00.0 & -0.042 & -0.054 & N/A & confusion (AX J1745.6-2901)\\
 $<\dots>$  &  & & &  &  \\
144 & SNR G094.0+01.0 & 93.974 & 1.026 & $<2.64$ & \\
145 & SNR G096.0+02.0 & 96.043 & 1.953 & $<2.79$ & \\
146 & SNR G106.3+02.7 & 106.273 & 2.705 & $<2.54$ & \\
147 & SNR G108.2$-$00.6 & 108.193 & -0.627 & $<2.27$ & \\
148 & SNR G109.1$-$01.0 & 109.142 & -1.015 & $2.63\pm0.73$ (3.60) & continuum (2E 2259.0+5836); $F_{78}~\textless~1\times10^{-5}~\textrm{ph~} \textrm{cm}^{-2}~\textrm{s}^{-1}$\\
149 & SNR G111.7$-$02.1 & 111.734 & -2.145 & $7.31\pm0.68$ (10.75) & Cas A\\
150 & SNR G113.0+00.2 & 114.088 & -0.213 & $<1.89$ & \\
 $<\dots>$  &  & & &  &  \\
294 & SNR G359.1$-$00.5 & -0.879 & -0.506 &  $<1.17$ & confusion (IGR J17446-2947) \\
\hline
\end{tabular}
\vspace{2mm}
\begin{tabular}{l}
\begin{minipage}{\textwidth}
$^1$  Catalogue is sorted by source name, which is in turn,
based on Galactic coordinates.  \\
$^2$  The upper limits are given at $3\sigma$ confidence level. In case of a source detection, the significance is shown in parenthesis. \\
$^3$  The spatial
confusion with a known hard X-ray source (shown in parenthesis) is indicated. SNRs dominated by the continuum flux are marked as well. If significance of the detection of such source is above $3\sigma$ level, an upper limit ($3\sigma$) in the 78.4 keV line is shown assuming a simple power-law continuum model.
\end{minipage}
\end{tabular}
\end{table*}
%=================================================================

\subsection{Historical supernovae and other reported sources of $^{44}$Ti}
\label{hist}

One of the main goals of this survey was to verify the $^{44}$Ti line
fluxes from historical SNRs reported in the literature and to provide
an independent estimate of the corresponding amount of synthesised
$^{44}$Ti with much larger statistics available thanks to the whole
{\it INTEGRAL} IBIS/ISGRI dataset. List of the young (with ages not
more than $\sim1000$ years) Galactic SNRs with known (or estimated)
dates of explosion and distances used in our study is presented in
Table \ref{tab:hist}.

To estimate the $^{44}$Ti flux in the most robust way we used the
simplest model for the continuum emission -- power-law.  This model is
usually assumed for fitting the continuum of SNRs in the hard X-ray
domain although it is known that synchrotron emission does exhibit
spectral steepening beyond the cutoff energy
\citep[e.g.][]{1998ApJ...493..375R,2007A&A...465..695Z}. To take into
account $^{44}$Ti emission lines we introduced to the model two
Gaussians with fixed positions at 67.9 and 78.4 keV, and intrinsic
widths fixed at $10^{-3}$ keV.  Additionally fluxes in the two lines
were tied as $F_{68}=0.93 F_{78}$.  For the fitting procedure we used
XSPEC v.12.8.1g.

%=================================================================
%=================================================================

%======================Table1===========================================
%\begin{landscape}
\begin{table*}
\noindent
\centering
\caption{List of the young (with ages not more than $\sim1000$ years) Galactic SNRs with known (or estimated) dates of explosion and distances.}\label{tab:hist}
\centering
\vspace{1mm}
\small{
  \begin{tabular}{c|c|c|c|c|c|c|c}
\hline\hline
Name   & $l$ & $b$ & Exposure, Ms$^1$ & SNR age, years & Distance, kpc & Type & References$^2$\\
\hline
G1.9+0.3 & 1.871 & 0.326 & 18.1 & $110\pm11$ & 8.5 &  Ia & 1, 2  \\
Kepler   & 4.526 & 6.820 & 11.2 & 410 & $3.9^{+1.4}_{-0.9}$ & Ia & 3, 4  \\
G21.5$-$0.9 & 21.501 & $-0.885$ & 5.1 & $870^{+200}_{-150}$ & $4.7\pm0.4$ & II & 5, 6  \\
Kes 75  & 29.702 & $-0.248$ & 4.8 & $\sim730$ & $6.3\pm1.2$ & II & 7, 8  \\
Cas A & 111.734 & $-2.145$ & 5.9 & 343 & $3.4^{+0.3}_{-0.1}$ & IIb  & 9, 10, 11  \\
Tycho & 120.087 & 1.423 & 6.8 & 442  & $4\pm1$ & Ia & 12, 13  \\
SN 1181 & 130.718 & 3.084 & 2.7 & 833 & $2.0\pm0.3$ & II & 14  \\
Crab & 184.557 & $-5.784$ & 3.8 & 960 & $2.0\pm0.5$ & II & 15  \\
G310.6$-$1.6  & 310.592 & $-1.593$ & 3.3 & $\sim1000$ & $7\pm2$ & II & 16  \\
SN 1006 & 327.572 & 14.566 & 2.5 & 1008  & $2.18\pm0.08$ & Ia & 17 \\
%Kes 73  & 27.386620 & $-0.006000$ &  & 1.3 kyr & $6.8\pm0.8$ &  & \\ Sanbonmatsu & Helfand 1992 (see Vink & Kuiper 2006, Tian & Leahy 2008
%CTB 37B  & 348.641506 & 0.434565 &  & $\sim880$  & $\sim8$ & II & 29, \\ hosting a ~1kyr old magnetar, Halpern & Gotthelf 2010
\hline
  \end{tabular}
\vspace{3mm}
\begin{minipage}{0.8\textwidth}
$^1$~Dead-time corrected effective exposure of the \textit{INTEGRAL}/IBIS telescope \\
$^2$~References: (1) \cite{2008ApJ...680L..41R}, (2) \cite{2011ApJ...737L..22C}, (3) \cite{2005AdSpR..35.1027S}, (4) \cite{2007ApJ...668L.135R}, (5) \cite{2008MNRAS.386.1411B}, (6) \cite{2006ApJ...637..456C}, (7) \cite{2000ApJ...542L..37G}, (8) \cite{2008A&A...480L..25L}, (9) \cite{2001AJ....122..297T}, (10) \cite{1995ApJ...440..706R}, (11) \cite{2008Sci...320.1195K}, (12) \cite{2010ApJ...725..894H}, (13) \cite{2008Natur.456..617K}, (14) \cite{2013A&A...560A..18K}, (15) \cite{1973PASP...85..579T}, (16) \cite{2010ApJ...716..663R}, (17) \cite{2003ApJ...585..324W}.
\end{minipage}}
\end{table*}
%\end{landscape}
%=========================================================================

\centerline{}
\noindent{\bf Cassiopeia A}

First detection of $^{44}$Ti line emission from Cas\,A has been
reported by CGRO/COMPTEL \citep{1994A&A...284L...1I}, with a flux of
$(3.4\pm0.9)\times10^{-5}$\fl \citep{2000AIPC..510...54S}.  Later,
this finding was confirmed by \textit{BeppoSAX} observations which
resulted in detection of the 67.9 and 78.4 keV lines
\citep{2001ApJ...560L..79V}. The line flux was dependent on the
continuum shape, being $(1.9\pm0.4)\times10^{-5}$\fl assuming a simple
power-law spectrum.

\cite{2006ApJ...647L..41R} using about 4.5 Ms of
\textit{INTEGRAL}/ISGRI total exposure time detected both 67.9 and
78.4 keV lines with a flux of $(2.5\pm0.3)\times10^{-5}$\fl that
corresponds to a synthesised $^{44}$Ti mass of
$1.6^{+0.6}_{-0.3}\times10^{-4}$ M$_{\odot}$. In a more recent work by
\cite{2015A&A...579A.124S} the \textit{INTEGRAL}/SPI data were
utilized. Measured fluxes in the 78.4 keV and 1157 keV lines are
$(2.1\pm0.4)\times10^{-5}$\fl and $(3.5\pm1.2)\times10^{-5}$\fl, which
corresponds to $(1.5\pm0.4)\times10^{-4}$ and
$(2.4\pm0.9)\times10^{-4}$ M$_{\odot}$ of $^{44}$Ti,
respectively. Authors also pointed out that the reason for a
significantly larger flux in the 1157 keV line possibly could be an
additional contribution to this line from nuclear de-excitation
following energetic particle collisions in the remnant and swept-up
material.

Using the high spatial resolution and sensitivity of the
\textit{NuSTAR} focusing high-energy X-ray telescope,
\cite{2014Natur.506..339G} were able to map the distribution of
$^{44}$Ti emission over the Cas A supernova remnant.  Their result
confirmed previous claims of a highly asymmetric explosion, needed to
explain the observed amount of synthesised $^{44}$Ti. Spectroscopic
studies also revealed a redshift of the 67.9 keV line by $0.47\pm0.21$
keV, with a flux of $(1.5\pm0.3)\times10^{-5}$\fl.

Our analysis confirmed the presence of a highly significant emission
component in the E7 energy band above the continuum described by a
simple power-law model with photon index of $2.9\pm0.1$. The obtained flux
in the 67.9 keV line is $F_{68}=(1.3\pm0.3)\times10^{-5}$
\phflux. Both continuum and line parameters are compatible within
$1-2\sigma$ with results from \cite{2006ApJ...647L..41R} based on
smaller sample of IBIS/ISGRI data. Also this values agree very well
with the recent {\it NuSTAR} result and most of previously published
measurements.

\centerline{}
\noindent{\bf G1.9+0.3}

One of the most promising sources of $^{44}$Ti emission in our sample
is G1.9+0.3. Based on the detected expansion between 1985 and 2007
\cite{2008ApJ...680L..41R} estimated its age around 100 years, making
this SNR the youngest one in the Galaxy at distance around 8.5 kpc
from the Sun. Using the \textit{Chandra} observations
\cite{2010ApJ...724L.161B} detected a 4.1 keV line which was
attributed to $^{44}$Sc fluorescence emission. The estimated mass of
synthesised $^{44}$Ti is $(1$--$7)\times10^{-5}$ M$_{\odot}$ that
corresponds to an expected flux in the 67.9 and 78.4 keV nuclear
de-excitation lines around $(0.3$--$2.3)\times10^{-5}$\fl.  Later,
the \textit{NuSTAR} observatory put a 95\% upper limit for the 67.9
keV line around $1.5\times10^{-5}$\fl for an assumed line width (1
sigma) of 4 keV \citep{2015ApJ...798...98Z}.

In our survey we did not detect a significant flux from G1.9+0.3 with
a corresponding $3\sigma$ upper limit of $9\times10^{-6}$
\phflux. This upper limit is more constraining than the one obtained
with {\it NuSTAR} but still compatible within the errors with the
reported 4.1 keV line flux.

\centerline{}
\noindent{\bf GRO\,J0852-4642 / RX\,J0852.0-4622 / G266.2-1.2 / Vela Jr}

COMPTEL discovered a source of $^{44}$Ti line emission from a
previously unknown Galactic SNR in the Vela region, named Vela Jr,
with a flux of $(3.8\pm0.7)\times10^{-5}$\fl in the 1.157 MeV line
\citep{1998Natur.396..142I}.  Later \cite{2000AIPC..510...54S} have
shown this detection to be much less significant due to technical
issues related to the background modeling and event selection. Doubts
about the reality of the $^{44}$Ti line detection were also raised by
the age and distance estimates pointed towards a more distant and
older SNR than originally thought
\citep[e.g.][]{2001ApJ...548..814S,2008ApJ...678L..35K}. At lower
energies some evidence of 4.1 keV line identified with $^{44}$Sc were
reported by \cite{2000PASJ...52..887T}, \cite{2005A&A...429..225I} and
\cite{2005ApJ...632..294B}. However, \cite{2001ApJ...548..814S} and
\cite{2009PASJ...61..275H} did not confirm these detections, making
future studies very important.

In our survey we did not detect the signal from the $^{44}$Ti
decay. The corresponding $3\sigma$ upper limit is $1.8\times10^{-5}$
\phflux. Thus the {\it INTEGRAL}/IBIS measurement excludes the
detection made by the COMPTEL experiment under assumption of a
point-like source, confirming results of similar analysis performed by
\cite{2006NewAR..50..540R}. Taking into account that SNR has a
diameter of 2\arcdeg, a more detailed study should be carried out in
order to properly derive a consistent upper limit as a function of the
source extent.

\centerline{}
\noindent{\bf Tycho's Supernova / SN\,1572}

The \textit{Swift}/BAT telescope discovered a significant emission
excess above the continuum level in the 60--85 keV energy band in the
spectrum of Tycho SNR \citep{2014ApJ...797L...6T}. It is the
first evidence for a detection of $^{44}$Ti decay emission in a Type
Ia SNR. The flux in the 78.4 keV line was measured to be
$(1.4\pm0.6)\times10^{-5}$\fl that corresponds to a $^{44}$Ti mass
around $10^{-4}$ M$_{\odot}$ depending on the distance to the SNR
\citep[see][]{2014ApJ...797L...6T}. Marginal detection (significance
level $\sim2.6\sigma$) of a bump feature in the 60--90 keV band was
reported also by \cite{2014ApJ...789..123W} using the
\textit{INTEGRAL}/IBIS telescope data. The corresponding $3\sigma$
upper limit for the $^{44}$Ti line emission of $1.5\times10^{-5}$\fl which
coincides with earlier result obtained by \cite{2006NewAR..50..540R}
using the \textit{INTEGRAL}/IBIS data as well. An indication of the spatial
co-location of the post-shock Ti with other iron-peak nuclei was found
using \textit{XMM-Newton} data \citep{2015ApJ...805..120M}.

However, the more recent \textit{NuSTAR} observations did not find
evidence for $^{44}$Ti emission in the spectrum of Tycho SNR with the
upper limit $F_{78}<7.5\times10^{-6}$ \phflux~(90\% confidence level),
ruling out the above-mentioned \textit{Swift}/BAT and
\textit{INTEGRAL}/IBIS detections \citep{2015ApJ...814..132L}. The
corresponding upper limit on the $^{44}$Ti yield was found to be
$M_{44} < 8.4 \times 10^{-5}$ M$_{\odot}$ assuming distance to the SNR
of 2.3 kpc.

To shed further light on the existence of the $^{44}$Ti emission in
the spectrum of Tycho SNR we have performed a fit to the measured
IBIS/ISGRI spectrum, as for the other young SNRs in our sample. With
the achieved sensitivity we can not confidently confirm the detection
made by \textit{Swift}/BAT \citep{2014ApJ...797L...6T}. Inclusion of
the $^{44}$Ti lines into the fitting model consisting of a simple
power-law  gives a flux in the 78.4
keV line $F_{78}=(5\pm3)\times10^{-6}$ \phflux, compatible with the
above-mentioned upper limit obtained with {\it NuSTAR}
\citep{2015ApJ...814..132L}.

%(with photon index of $2.2\pm0.4$)

\centerline{}
\noindent{\bf Kepler}

Detection of the $^{44}$Ti emission from the Kepler SNR has not been
claimed so far. $3\sigma$ upper limit in 1.157 MeV line obtained in
the COMPTEL Galactic survey is $1.8\times10^{-5}$
\phflux\ \citep{1997A&A...324..683D,1999ApL&C..38..383I}. We did not
detect a significant flux from the source in any working energy bands
(see Tab. \ref{tab:ebands}) either. Upper limit obtained in our survey
is $6.3\times10^{-6}$ \phflux ($3\sigma$) per line.

\centerline{}
\noindent{\bf SN\,1987A} 

The only known extragalactic source of $^{44}$Ti decay emission is
SN\,1987A in the Large Magellanic Cloud where the \textit{INTEGRAL}/IBIS
telescope detected significant flux in the 67.9 and 78.4 keV lines
\citep{2012Natur.490..373G}.  The combined flux in both lines was
measured to be $(1.5\pm0.4)\times10^{-5}$\fl that can be translated into
a $^{44}$Ti mass of $(3.1\pm0.8)\times10^{-4}$ M$_{\odot}$. Later the
\textit{NuSTAR} telescope confirmed the existence of both lines, but
with somewhat lower flux corresponding to a $^{44}$Ti yield of
$(1.5\pm0.3)\times10^{-4}$ M$_{\odot}$
\citep{2015Sci...348..670B}. 

The {\it INTEGRAL} survey described above includes additional data
compared to \cite{2012Natur.490..373G} and leads to a slightly
smaller line flux of $\sim1\times10^{-5}$\fl corresponding to $^{44}$Ti mass
of $\sim2\times10^{-4}$ M$_{\odot}$ which is closer to the value
obtained by {\it NuSTAR}. Because SN\,1987A is located in another
galaxy (LMC) a detailed analysis will be published in a separate
paper (Grebenev et al., in preparation).

\centerline{}
\noindent{\bf Per OB2} 

In addition to the aforementioned sources one can point another
potential source of $^{44}$Ti emission located in the region of
Perseus OB2 association.  Some hints of a weak signal (at significance
level around $\sim3\sigma$) in the 1.157 MeV line from this region was
reported by \cite{1997A&A...324..683D} using the COMPTEL data.  We did
not detect any significant point-like source in the vicinity of Per OB2
with $3\sigma$ upper limit $\sim1.5$ mCrab ($3.3\times10^{-5}$
\phflux).

\section{Summary}

In this work we performed a systematic search for $^{44}$Ti line
emission at 67.9 and 78.7 keV with the IBIS/ISGRI instrument aboard
the {\it INTEGRAL} observatory. The long exposure collected by the
mission in the Galactic Plane allowed us to put strong limits on the
$^{44}$Ti line emission and improve the previous constraints made with
COMPTEL experiment in the 1.157 MeV $^{44}$Ti line emission by a
factor of 2--5.

%=================================================================
\begin{figure}
\includegraphics[width=\columnwidth,bb=15 175 545 585, clip]{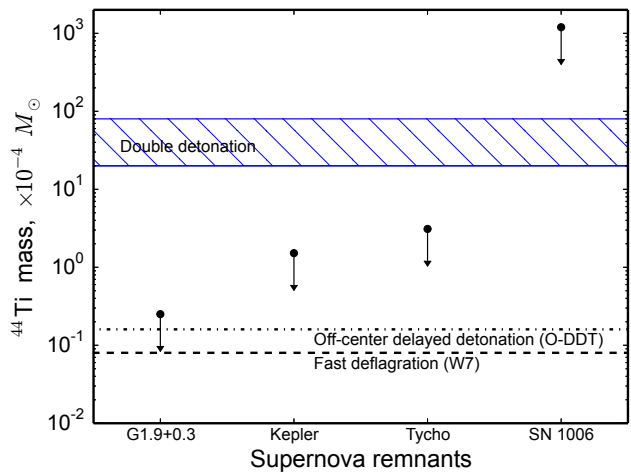}

\caption{Arrows show $3\sigma$ upper limits on the $^{44}$Ti yield
  for Type Ia historical SNe obtained in this work. Ages and
  distances for each SNR are taken from Table \ref{tab:hist}. The
  $^{44}$Ti mass predictions from different explosion models are shown
  as well. }\label{fig:res}

\end{figure}
%=================================================================

Sensitivity of the current survey is governed by photons count
statistics only which allowed us to set up the lowest possible
detection threshold while not flooding the catalog of candidate
sources by false detections. Expecting only one statistical
fluctuation above $4.7\sigma$, we detected a $^{44}$Ti source
candidate IGR~J18406+0451 with S/N=$5\sigma$. However, deep {\it
  XMM-Newton} observation does not show the presence of
a bright source in this region, indicating that IGR\,J18406+0451 is
either a false detection or its emission in soft X-rays is strongly
absorbed (either locally or due to the ISM gas along the line of sight).

Among historical SNRs only Cas A shows significant detection with a
flux $F_{68}=(1.3\pm0.3)\times10^{-5}$ \phflux\ in agreement with most
of previously published measurements. This flux corresponds to
$(1.1\pm0.3)\times10^{-4}$ M$_{\odot}$ of $^{44}$Ti that is at the
upper boundary of the theoretical predictions for core-collapse SNe
\citep[see, e.g.,][]{1996ApJ...464..332T}.

We put strong upper limits on the detection of $^{44}$Ti emission in
other historical SNRs and potential $^{44}$Ti sources mentioned in the
literature: Vela Jr, Tycho (SN1572), Per OB2 and
G1.9+0.3. Regarding the latter, we derived a more
constraining upper limit on the $^{44}$Ti line emission than that
obtained with {\it NuSTAR} \citep{2015ApJ...798...98Z} by a factor of
few.

Predicted $^{44}$Ti yield for Type Ia SNe is scattered from
$\sim8\times10^{-6}$ M$_{\odot}$ in the classical deflagration model
\citep[W7 in][]{1984ApJ...286..644N,2010ApJ...712..624M} to
$few\times10^{-3}$ M$_{\odot}$ in the so-called double detonation
sub-Chandrasekhar model, when the explosion is triggered by an
explosion at the surface of the white dwarf \citep[models 1-4
  in][]{2010A&A...514A..53F}.

Figure \ref{fig:res} shows the comparison of the $3\sigma$ upper
limits on the $^{44}$Ti yield from Type Ia SNe obtained in this work
with predictions of different models. Such extreme models as double
detonation \citep{2010A&A...514A..53F} or helium deflagration
\citep{2011ApJ...734...38W}, predicting $>10^{-3}$ M$_{\odot}$ of $^{44}$Ti,
are clearly ruled out by the observational data. At the same time broad
range of models predicting $^{44}$Ti yields below $\sim10^{-5}$
M$_{\odot}$ is consistent with our results. For the illustration
purposes in Fig. \ref{fig:res} the fast deflagration W7 model
\citep{1984ApJ...286..644N} and delayed detonation model which follows an
extremely off-center deflagration \citep[O-DDT model
  in][]{2010ApJ...712..624M} are shown by dashed and dash-dot lines,
respectively.

Finally we put strong upper limits for all Galactic SNRs reported in
the catalog by \cite{green2014}. The obtained upper limits can be used
to estimate the exposure times needed to detect $^{44}$Ti emission
from any known SNR using existing and prospective X- and gamma-ray
telescopes. Moreover, more or less homogeneous coverage of the
Galactic Plane with {\it INTEGRAL} (see Fig. \ref{fig:sen}) permits to
estimate an upper limit on the titanium emission from any potentially
interesting place in the Galaxy: starburst regions, high-absorption
regions such as the spiral arm tangents, and to revisit the issue
about the main sources of Galactic $^{44}$Ca
\citep{2006A&A...450.1037T,2013ApJ...775...52D}. In particular,
  well known giant molecular cloud Sagittarius B2 (Sgr B2) located
  about 120 pc from the center of the Milky Way emits not more than
  $5.1\times10^{-6}$ \fl\ ($3\sigma$ upper limit) per $^{44}$Ti line;
  the Orion Molecular Cloud Complex containing many bright nebulae,
  dark clouds, and young stars has a $3\sigma$ upper limit of
  $1.5\times10^{-5}$ \fl per line; a $3\sigma$ upper limit for the
  flux from the Norma and Sagittarius spiral arms tangents is
  $1.0\times10^{-5}$ \fl\ per line. Among extragalactic objects one of
  the most interesting is the Tarantula Nebula (known as 30 Doradus)
  in the Large Magellanic Cloud. This most active starburst region in
  the Local Group emits less than $0.9\times10^{-6}$ \fl.

\centerline{}

%\onecolumn
%\include{green}
%\twocolumn

\section*{Acknowledgments}

The work was supported by Russian Science Foundation (grant
14-22-00271). We thank the anonymous referee whose suggestions helped
improve and clarify the manuscript. The results of this work are based
on observations of the INTEGRAL observatory, an ESA project with the
participation of Denmark, France, Germany, Italy, Switzerland, Spain,
the Czech Republic, Poland, Russia and the United States. Results are
also partly based on observations obtained with XMM-Newton, an ESA
science mission with instruments and contributions directly funded by
ESA Member States and NASA. We are grateful to the XMM-Newton team for
the acceptance and execution of our DDT request. Authors thank Max
Planck Institute for Astrophysics for computational support.

\onecolumn
\begin{minipage}{0.8\textwidth}
{\bf Table 2.} The complete catalogue of Galactic SNRs (Green, 2014) with measured fluxes or
$3\sigma$ upper limits in the $64.6-82.2$~keV band. This catalogue is only
available in the online version of the paper.
\end{minipage}
\begin{center}
\begin{longtable}{|c|l|r|r|c|l|c|c|c|c|}
%\label{tab:green}

\hline 
\multicolumn{1}{|c|}{No.} &
\multicolumn{1}{c|}{Name\footnotemark[1]} & 
\multicolumn{1}{c|}{l$^{II}$} & 
\multicolumn{1}{c|}{b$^{II}$} &
\multicolumn{1}{c|}{Flux$_{\rm 64.6-82.2~keV}$\footnotemark[2]} &
\multicolumn{1}{c|}{Notes\footnotemark[3]} \\

\multicolumn{1}{|c|}{} &
\multicolumn{1}{c|}{} &
\multicolumn{1}{c|}{deg} &
\multicolumn{1}{c|}{deg} &
\multicolumn{1}{c|}{$10^{-12} \textrm{erg~} \textrm{cm}^{-2}~\textrm{s}^{-1}$} &
\multicolumn{1}{c|}{} \\ \hline 
\endfirsthead

\multicolumn{6}{c}%
{{\bfseries Table 2  -- continued from previous page}} \\
\hline 
\multicolumn{1}{|c|}{No.} &
\multicolumn{1}{c|}{Name\footnotemark[1]} &          
\multicolumn{1}{c|}{l$^{II}$} &                     
\multicolumn{1}{c|}{b$^{II}$} &
\multicolumn{1}{c|}{Flux$_{\rm 64.6-82.2~keV}$\footnotemark[2]} &
\multicolumn{1}{c|}{Notes\footnotemark[3]} \\

\multicolumn{1}{|c|}{} &
\multicolumn{1}{c|}{} &
\multicolumn{1}{c|}{deg} &
\multicolumn{1}{c|}{deg} &
\multicolumn{1}{c|}{$10^{-12} \textrm{erg~} \textrm{cm}^{-2}~\textrm{s}^{-1}$} &
\multicolumn{1}{c|}{} \\ \hline 
\endhead

\hline \multicolumn{6}{|r|}{{Continued on next page}} \\ \hline
\endfoot

\hline \hline
\endlastfoot

001 & SNR G000.0+00.0 & -0.042 & -0.054 & N/A & confusion (AX J1745.6-2901)\\
002 & SNR G000.3+00.0 & 0.330 & 0.040 & N/A & confusion (1E 1743.1-2843)\\
003 & SNR G000.9+00.1 & 0.869 & 0.084 & N/A & confusion (IGR J17464-2811) \\
004 & SNR G001.0$-$00.1 & 1.000 & -0.133 & N/A & confusion (IGR J17475-2822) \\
    &                   &  & & & confusion (IGR J17497-2821, IGR J17464-2811) \\
005 & SNR G001.4$-$00.1 & 1.460 & -0.154 & N/A & confusion (AX J1749.1-2733) \\
006 & SNR G001.9+00.3 & 1.871 & 0.326 & $<1.11$  & G1.9+0.3 \\
007 & SNR G003.7$-$00.2 & 3.781 & -0.280 & $<1.16$ & \\
008 & SNR G003.8+00.3 & 3.810 & 0.395 & $<1.16$ & \\
009 & SNR G004.2$-$03.5 & 4.210 & -3.506 & $<1.22$ & \\
010 & SNR G004.5+06.8 & 4.526 & 6.820 & $<1.50$ & Kepler SN1604 \\
011 & SNR G004.8+06.2 & 4.798 & 6.243 & $<1.46$ & \\
012 & SNR G005.2$-$02.6 & 5.197 & -2.600 & $<1.23$ & \\
013 & SNR G005.4$-$01.2 & 5.348 & -1.133 & N/A & confusion (GX 5-1)\\
014 & SNR G005.5+00.3 & 5.551 & 0.322 & $<1.23$ & \\
015 & SNR G005.9+03.1 & 5.905 & 3.130 & $<1.33$ & \\
016 & SNR G006.1+00.5 & 6.104 & 0.532 & $<1.26$ & \\
017 & SNR G006.1+01.2 & 6.096 & 1.209 & $<1.26$ & \\
018 & SNR G006.4+04.0 & 6.416 & 4.026 & $<1.40$ & \\
019 & SNR G006.4$-$00.1 & 6.435 & -0.076 & $<1.26$ & \\
020 & SNR G006.5$-$00.4 & 6.510 & -0.478 & $<1.26$ & \\
021 & SNR G007.0$-$00.1 & 7.050 & -0.078 & $<1.29$ & \\
022 & SNR G007.2+00.2 & 7.200 & 0.197 & $<1.30$ & \\
023 & SNR G007.7$-$03.7 & 7.754 & -3.771 & $<1.40$ & \\
024 & SNR G008.3$-$00.0 & 8.304 & -0.095 & $<1.38$ & \\
025 & SNR G008.7$-$00.1 & 8.744 & -0.096 & $<1.41$ & \\
026 & SNR G008.7$-$05.0 & 8.712 & -5.011 & $<1.56$ & \\
027 & SNR G008.9+00.4 & 8.903 & 0.403 & $<1.42$ & \\
028 & SNR G009.7$-$00.0 & 9.699 & -0.062 & $<1.71$ & confusion (SGR 1806-20) \\
029 & SNR G009.8+00.6 & 9.749 & 0.566 & $<1.52$ & \\
030 & SNR G009.9$-$00.8 & 9.958 & -0.805 & $<1.52$ & \\
031 & SNR G010.5$-$00.0 & 10.599 & -0.036 & $<1.56$ & \\
032 & SNR G011.0$-$00.0 & 11.027 & -0.051 & $<1.78$ & confusion (XTE J1810-189)\\
033 & SNR G011.1+00.1 & 11.184 & 0.112 & $<1.78$ & confusion (XTE J1810-189) \\
034 & SNR G011.1$-$00.7 & 11.143 & -0.713 &  $<1.78$ & confusion (PSR J1811-1925)\\
035 & SNR G011.1$-$01.0 & 11.171 & -1.042 & $<1.64$ & G11.2-0.3 SN386 \\
036 & SNR G011.2$-$00.3 & 11.184 & -0.337 & $<1.78$ & continuum (PSR J1811-1925) \\
037 & SNR G011.4$-$00.1 & 11.400 & -0.038 &  $<1.78$ & confusion (XTE J1810-189) \\
038 & SNR G011.8$-$00.2 & 11.893 & -0.208 & $<1.67$ & confusion (SNR G012.0$-$00.1) \\
039 & SNR G012.0$-$00.1 & 11.969 & -0.104 & $<1.67$ & confusion (SNR G011.8$-$00.2)\\
040 & SNR G012.2+00.3 & 12.260 & 0.300 & $<1.69$ & \\
041 & SNR G012.5+00.2 & 12.588 & 0.222 & $<1.91$ & confusion (IGR J18135-1751) \\
042 & SNR G012.7$-$00.0 & 12.727 & 0.004 & $<1.91$ & confusion (IGR J18135-1751)\\
043 & SNR G012.8$-$00.0 & 12.834 & -0.019 &  $<1.91$& confusion (IGR J18135-1751) \\
044 & SNR G013.3$-$01.3 & 13.319 & -1.302 & $<1.75$ & \\
045 & SNR G013.5+00.2 & 13.446 & 0.147 & $<1.86$ & confusion (GX 13+1)\\
046 & SNR G014.1$-$00.1 & 14.177 & -0.118 & $<1.91$ & confusion (SNR G014.3+00.1) \\
047 & SNR G014.3+00.1 & 14.303 & 0.141 & $<1.91$ & confusion (SNR G014.1$-$00.1) \\
048 & SNR G015.1$-$01.6 & 15.109 & -1.612 & $<1.88$ & \\
049 & SNR G015.4+00.1 & 15.419 & 0.179 & $<1.86$ & confusion (IGR J18175-1530) \\
050 & SNR G015.9+00.2 & 15.881 & 0.199 & $<1.88$ & \\
051 & SNR G016.0$-$00.5 & 16.054 & -0.548 & $<1.91$ & \\
052 & SNR G016.2$-$02.7 & 16.125 & -2.614 & $<1.95$ & \\
053 & SNR G016.4$-$00.5 & 16.412 & -0.549 & $<1.92$ & \\
054 & SNR G016.7+00.1 & 16.734 & 0.089 & $<1.89$ & confusion (AX J1820.5-1434)\\
055 & SNR G017.0$-$00.0 & 17.026 & -0.035 & $<1.89$ & confusion (SNR G016.7+00.1, IGR J18219-1347)\\
056 & SNR G017.4$-$00.1 & 17.485 & -0.116 & $<1.89$ & confusion (IGR J18219-1347)\\
057 & SNR G017.4$-$02.3 & 17.388 & -2.297 & $<2.00$ & \\
058 & SNR G017.8$-$02.6 & 17.795 & -2.608 & $<2.03$ & \\
059 & SNR G018.1$-$00.1 & 18.163 & -0.151 & $<2.03$ & \\
060 & SNR G018.6$-$00.2 & 18.627 & -0.278 & $<1.95$ & \\
061 & SNR G018.8+00.3 & 18.802 & 0.353 & $<1.95$ & \\
062 & SNR G018.9$-$01.1 & 18.952 & -1.185 & $<2.04$ & \\
063 & SNR G019.1+00.2 & 19.148 & 0.269 & $<1.95$ & \\
064 & SNR G020.0$-$00.2 & 19.983 & -0.171 & $<1.98$ & \\
065 & SNR G020.4+00.1 & 20.469 & 0.158 & $<1.98$ & \\
066 & SNR G021.0$-$00.4 & 21.043 & -0.470 & $<2.00$ & \\
067 & SNR G021.5$-$00.1 & 21.562 & -0.097 & $<2.00$ & confusion (AX J183039-1002) \\
068 & SNR G021.5$-$00.9 & 21.487 & -0.890 & $7.8\pm0.67$ (11.64) & continuum SNR 021.5-00.9; $F_{78}~\textless~9\times10^{-6}~\textrm{ph~} \textrm{cm}^{-2}~\textrm{s}^{-1}$\\
069 & SNR G021.6$-$00.8 & 21.648 & -0.839 & $<2.01$ & confusion (SNR 021.5-00.9) \\
070 & SNR G021.8$-$00.6 & 21.795 & -0.508 & $<2.11$ & \\
071 & SNR G022.7$-$00.2 & 22.665 & -0.194 & $<2.01$ & \\
072 & SNR G023.3$-$00.3 & 23.206 & -0.331 & $<2.03$ & \\
073 & SNR G023.6+00.3 & 23.530 & 0.311 & $<0.52$ & confusion (IGR J18325-0756) \\
074 & SNR G024.7+00.6 & 24.663 & 0.588 & $<2.04$ & \\
075 & SNR G024.7$-$00.6 & 24.782 & -0.621 & $<2.07$ & \\
076 & SNR G025.1$-$02.3 & 25.097 & -2.255 & $<2.15$ & \\
077 & SNR G027.4+00.0 & 27.389 & -0.004 & $2.94\pm0.67$ (4.39) & continuum (IGR J18325-0756); $F_{78}~\textless~1\times10^{-5}~\textrm{ph~} \textrm{cm}^{-2}~\textrm{s}^{-1}$\\
078 & SNR G027.8+00.6 & 27.694 & 0.569 & $<2.08$ & \\
079 & SNR G028.6$-$00.1 & 28.619 & -0.100 & $<2.10$ & \\
080 & SNR G028.8+01.5 & 28.917 & 1.433 & $<2.10$ & \\
081 & SNR G029.6+00.1 & 29.558 & 0.115 & $<2.22$ & \\
082 & SNR G029.7$-$00.3 & 29.705 & -0.244 & $4.80\pm0.69$ (6.96)  & continuum (PSR J1846-0258); $F_{78}~\textless~1\times10^{-5}~\textrm{ph~} \textrm{cm}^{-2}~\textrm{s}^{-1}$ \\
083 & SNR G030.7+01.0 & 30.719 & 0.955 & $<2.08$ & \\
084 & SNR G030.7$-$02.0 & 30.689 & -1.984 & $<2.28$ & \\
085 & SNR G031.5$-$00.6 & 31.551 & -0.631 & $<2.07$ & \\
086 & SNR G031.9+00.0 & 31.886 & 0.032 & $<2.28$ & confusion (IGR J18486-0047) \\
087 & SNR G032.0$-$04.9 & 31.916 & -4.606 & $<2.28$ & \\
088 & SNR G032.1$-$00.9 & 32.120 & -0.901 & $<2.17$ & confusion (IGR J18538-0102) \\
089 & SNR G032.4+00.1 & 32.407 & 0.111 & $<2.17$ & \\
090 & SNR G032.8$-$00.1 & 32.811 & -0.056 & $<2.03$ & \\
091 & SNR G033.2$-$00.6 & 33.175 & -0.548 & $<2.01$ & \\
092 & SNR G033.6+00.1 & 33.695 & 0.009 & $<2.00$ & \\
093 & SNR G034.7$-$00.4 & 34.668 & -0.392 & $<1.97$ & \\
094 & SNR G035.6$-$00.4 & 35.643 & -0.430 & $<1.92$ & \\
095 & SNR G036.6+02.6 & 36.582 & 2.602 & $<1.95$ & \\
096 & SNR G036.6$-$00.7 & 36.585 & -0.695 & $<1.89$ & \\
097 & SNR G038.7$-$01.3 & 38.644 & -1.341 & $<1.83$ & \\
098 & SNR G039.2$-$00.3 & 39.243 & -0.321 & $<1.81$ & \\
099 & SNR G039.7$-$02.0 & 39.694 & -2.387 & $<1.91$ & confusion (SS 433)\\
100 & SNR G040.5$-$00.5 & 40.522 & -0.509 & $<1.83$ & \\
101 & SNR G041.1$-$00.3 & 41.115 & -0.314 & $<1.81$ & \\
102 & SNR G041.5+00.4 & 41.480 & 0.359 & $<1.83$ & \\
103 & SNR G042.0$-$00.1 & 41.953 & -0.047 & $<1.81$ & \\
104 & SNR G042.8+00.6 & 42.820 & 0.635 &$<1.91$ & confusion (SGR 1900+14) \\
105 & SNR G043.3$-$00.2 & 43.267 & -0.190 & $<1.91$ & confusion (IGR J19108+0917) \\
106 & SNR G043.9+01.6 & 43.908 & 1.614 & $<1.89$ & \\
107 & SNR G045.7$-$00.4 & 45.686 & -0.390 & $<2.06$ & confusion (GRS 1915+105)\\
108 & SNR G046.8$-$00.3 & 46.771 & -0.302 & $<1.94$ & \\
109 & SNR G049.2$-$00.7 & 49.140 & -0.603 & $<2.08$ & \\
110 & SNR G053.6$-$02.2 & 53.626 & -2.261 & $<2.59$ & \\
111 & SNR G054.1+00.3 & 54.094 & 0.260 & $<2.64$ & \\
112 & SNR G054.4$-$00.3 & 54.474 & -0.291 & $<2.67$ & \\
113 & SNR G055.0+00.3 & 55.110 & 0.419 & $<2.79$ & \\
114 & SNR G055.7+03.4 & 55.599 & 3.515 & $<3.03$ & \\
115 & SNR G057.2+00.8 & 57.300 & 0.833 & $<2.97$ & \\
116 & SNR G059.5+00.1 & 59.579 & 0.116 & $<3.03$ & \\
117 & SNR G059.8+01.2 & 59.807 & 1.200 & $<3.03$ & \\
118 & SNR G063.7+01.1 & 63.786 & 1.165 & $<2.82$ & \\
119 & SNR G064.5+00.9 & 64.517 & 0.942 & $<2.76$ & \\
120 & SNR G065.1+00.6 & 65.268 & 0.302 & $<2.71$ & \\
121 & SNR G065.3+05.7 & 65.179 & 5.661 & $<2.98$ & \\
122 & SNR G065.7+01.2 & 65.716 & 1.208 & $1.92\pm0.89$ (2.16) & \\
123 & SNR G065.8$-$00.5 & 65.844 & -0.546 & $<2.71$ & \\
124 & SNR G066.0$-$00.0 & 66.027 & -0.048 & $<2.67$ & \\
125 & SNR G067.6+00.9 & 67.581 & 0.924 & $<2.55$ & \\
126 & SNR G067.7+01.8 & 67.738 & 1.823 & $<2.55$ & \\
127 & SNR G067.8+00.5 & 67.806 & 0.495 & $<2.55$ & \\
128 & SNR G068.6$-$01.2 & 68.603 & -1.204 & $<2.56$ & \\
129 & SNR G069.0+02.7 & 68.837 & 2.778 & $2.02\pm0.84$ (2.41) & \\
130 & SNR G069.7+01.0 & 69.690 & 1.000 & $<2.43$ & \\
131 & SNR G073.9+00.9 & 73.912 & 0.883 & $<2.25$ & \\
132 & SNR G074.0$-$08.5 & 73.982 & -8.564 & $<3.36$ & \\
133 & SNR G074.9+01.2 & 74.942 & 1.141 & $<2.28$ & confusion (IGR J20159+3713)\\
134 & SNR G076.9+01.0 & 76.895 & 0.972 & $<2.19$ & \\
135 & SNR G078.2+02.1 & 78.143 & 2.186 & $<2.21$ & \\
136 & SNR G082.2+05.3 & 82.150 & 5.316 & $<2.56$ & \\
137 & SNR G083.0$-$00.3 & 83.002 & -0.272 & $<2.28$ & \\
138 & SNR G084.2$-$00.8 & 84.195 & -0.807 & $<2.33$ & \\
139 & SNR G085.4+00.7 & 85.366 & 0.783 & $<2.37$ & \\
140 & SNR G085.9$-$00.6 & 85.906 & -0.607 & $<2.38$ & \\
141 & SNR G089.0+04.7 & 88.838 & 4.794 & $<2.73$ & \\
142 & SNR G093.3+06.9 & 93.278 & 6.906 & $<3.17$ & \\
143 & SNR G093.7$-$00.2 & 93.754 & -0.218 & $<2.62$ & \\
144 & SNR G094.0+01.0 & 93.974 & 1.026 & $<2.64$ & \\
145 & SNR G096.0+02.0 & 96.043 & 1.953 & $<2.79$ & \\
146 & SNR G106.3+02.7 & 106.273 & 2.705 & $<2.54$ & \\
147 & SNR G108.2$-$00.6 & 108.193 & -0.627 & $<2.27$ & \\
148 & SNR G109.1$-$01.0 & 109.142 & -1.015 & $2.63\pm0.73$ (3.60) & continuum (2E 2259.0+5836); $F_{78}~\textless~1\times10^{-5}~\textrm{ph~} \textrm{cm}^{-2}~\textrm{s}^{-1}$\\
149 & SNR G111.7$-$02.1 & 111.734 & -2.145 & $7.31\pm0.68$ (10.75) & Cas A\\
150 & SNR G113.0+00.2 & 114.088 & -0.213 & $<1.89$ & \\
151 & SNR G114.3+00.3 & 114.292 & 0.300 & $<1.88$ & \\
152 & SNR G116.5+01.1 & 116.485 & 1.104 & $<1.84$ & \\
153 & SNR G116.9+00.2 & 116.926 & 0.173 & $<1.84$ & \\
154 & SNR G119.5+10.2 & 119.579 & 10.170 & $<3.18$ & \\
155 & SNR G120.1+01.4 & 120.087 & 1.423 & $2.20\pm0.63$ (3.50) & continuum (Tycho SNR); $F_{78}~\textless~9\times10^{-6}~\textrm{ph~} \textrm{cm}^{-2}~\textrm{s}^{-1}$ \\
156 & SNR G126.2+01.6 & 126.245 & 1.575 & $<2.27$ & \\
157 & SNR G127.1+00.5 & 127.081 & 0.593 & $<2.34$ & \\
158 & SNR G130.7+03.1 & 130.728 & 3.075 & $<2.94$ & \\
159 & SNR G132.7+01.3 & 132.621 & 1.512 & $<3.13$ & \\
160 & SNR G152.4$-$02.1 & 152.560 & -2.052 & $<3.41$ & \\
161 & SNR G156.2+05.7 & 156.115 & 5.663 & $<3.78$ & \\
162 & SNR G159.6+07.3 & 159.598 & 7.300 & $<4.20$ & \\
163 & SNR G160.9+02.6 & 160.436 & 2.787 & $<3.70$ & \\
164 & SNR G166.0+04.3 & 166.116 & 4.278 & $<4.75$ & \\
165 & SNR G178.2$-$04.2 & 179.411 & -2.361 & $<3.22$ & \\
166 & SNR G179.0+02.6 & 179.052 & 2.597 & $<4.00$ & \\
167 & SNR G180.0$-$01.7 & -179.830 & -1.819 & $<3.18$ & \\
168 & SNR G182.4+04.3 & -177.583 & 4.299 & $<3.99$ & \\
169 & SNR G184.6$-$05.8 & -175.446 & -5.786 & N/A & continuum (Crab Nebula) \\
170 & SNR G189.1+03.0 & -170.967 & 2.978 & $<3.70$ & \\
171 & SNR G190.9$-$02.2 & -169.027 & -2.128 & $<3.27$ & \\
172 & SNR G192.8$-$01.1 & -167.234 & -1.109 & $<3.62$ & \\
173 & SNR G205.5+00.5 & -154.268 & 0.209 & $<6.60$ & \\
174 & SNR G206.9+02.3 & -153.112 & 2.315 & $<7.06$ & \\
175 & SNR G213.0$-$00.6 & -146.692 & -0.363 & $<5.49$ & \\
176 & SNR G260.4$-$03.4 & -99.600 & -3.439 & $<2.22$ & \\
177 & SNR G261.9+05.5 & -98.052 & 5.478 & $<2.75$ & confusion (1RXS J090431.1-382920)\\
178 & SNR G263.9$-$03.3 & -96.061 & -3.368 & $<2.07$ & \\
179 & SNR G266.2$-$01.2 & -93.741 & -1.220 & $<2.04$ & Vela Jr\\
180 & SNR G272.2$-$03.2 & -87.783 & -3.181 & $<2.46$ & \\
181 & SNR G279.0+01.1 & -81.368 & 1.219 & $<3.01$ & \\
182 & SNR G284.3$-$01.8 & -75.691 & -1.783 & $<2.69$ & \\
183 & SNR G286.5$-$01.2 & -73.434 & -1.214 & $<2.56$ & \\
184 & SNR G289.7$-$00.3 & -70.315 & -0.293 & $<2.49$ & \\
185 & SNR G290.1$-$00.8 & -69.851 & -0.780 & $<2.50$ & \\
186 & SNR G291.0$-$00.1 & -68.977 & -0.084 & $<2.49$ & \\
187 & SNR G292.0+01.8 & -67.970 & 1.755 & $<2.49$ & \\
188 & SNR G292.2$-$00.5 & -67.837 & -0.536 & $<2.61$ & \\
189 & SNR G293.8+00.6 & -66.230 & 0.605 & $<2.50$ & \\
190 & SNR G294.1$-$00.0 & -65.884 & -0.057 & $<2.52$ & \\
191 & SNR G296.1$-$00.5 & -63.947 & -0.503 & $<2.55$ & \\
192 & SNR G296.5+10.0 & -63.512 & 9.932 & $<3.57$ & \\
193 & SNR G296.7$-$00.9 & -63.339 & -0.945 & $<2.58$ & \\
194 & SNR G296.8$-$00.3 & -63.119 & -0.335 & $<2.56$ & \\
195 & SNR G298.5$-$00.3 & -61.472 & -0.327 & $<2.54$ & \\
196 & SNR G298.6$-$00.0 & -61.394 & -0.062 & $<2.54$ & \\
197 & SNR G299.2$-$02.9 & -60.815 & -2.892 & $<2.67$ & \\
198 & SNR G299.6$-$00.5 & -60.413 & -0.471 & $<2.54$ & \\
199 & SNR G301.4$-$01.0 & -58.559 & -0.984 & $<2.50$ & \\
200 & SNR G302.3+00.7 & -57.713 & 0.731 & $<2.46$ & \\
201 & SNR G304.6+00.1 & -55.401 & 0.125 & $<2.42$ & \\
202 & SNR G306.3$-$00.9 & -53.693 & -0.894 & $<2.42$ & \\
203 & SNR G308.1$-$00.7 & -51.867 & -0.662 & $<2.42$ & \\
204 & SNR G308.4$-$01.4 & -28.346 & -23.068 & $<5.80$ & \\
205 & SNR G308.8$-$00.1 & -51.189 & -0.096 & $<2.42$ & \\
206 & SNR G309.2$-$00.6 & -50.843 & -0.696 & $<2.43$ & \\
207 & SNR G309.8+00.0 & -50.215 & 0.000 & $<2.44$ & \\
208 & SNR G310.6$-$00.3 & -49.380 & -0.276 & $<2.46$ & \\
209 & SNR G310.6$-$01.6 & -49.410 & -1.597 & $3.51\pm0.83$ (4.23)& continuum (IGR J14003-6326); $F_{78}~\textless~1\times10^{-5}~\textrm{ph~} \textrm{cm}^{-2}~\textrm{s}^{-1}$\\
210 & SNR G310.8$-$00.4 & -49.189 & -0.465 & $<2.47$ & \\
211 & SNR G311.5$-$00.3 & -48.469 & -0.340 & $<2.49$ & \\
212 & SNR G312.4$-$00.4 & -47.571 & -0.375 & $<2.52$ & \\
213 & SNR G312.5$-$03.0 & -47.512 & -3.003 & $<2.59$ & \\
214 & SNR G315.1+02.7 & -44.914 & 2.826 & $2.26\pm0.89$ (2.53) & \\
215 & SNR G315.4$-$00.3 & -44.587 & -0.285 & $<2.62$ & \\
216 & SNR G315.4$-$02.3 & -44.583 & -2.364 & $<2.69$ & \\
217 & SNR G315.9$-$00.0 & -44.138 & -0.025 & $<2.64$ & \\
218 & SNR G316.3$-$00.0 & -43.712 & -0.013 & $<2.66$ & \\
219 & SNR G317.3$-$00.2 & -42.686 & -0.238 & $<2.67$ & \\
220 & SNR G318.2+00.1 & -41.785 & 0.094 & $<2.69$ & \\
221 & SNR G318.9+00.4 & -41.094 & 0.391 & $<2.70$ & \\
222 & SNR G320.4$-$01.2 & -39.615 & -1.198 & $27.04\pm0.91$ (29.71) & continuum (PSR 1509-58); $F_{78}~\textless~1\times10^{-5}~\textrm{ph~} \textrm{cm}^{-2}~\textrm{s}^{-1}$ \\
223 & SNR G320.6$-$01.6 & -39.322 & -1.536 & $<2.75$ & \\
224 & SNR G321.9$-$00.3 & -38.099 & -0.298 & $<2.79$ & \\
225 & SNR G321.9$-$01.1 & -38.109 & -1.066 & $<2.69$ & \\
226 & SNR G322.1+00.0 & -37.866 & 0.028 & $<2.79$ & confusion (4U 1516-569 ) \\
227 & SNR G322.5$-$00.1 & -37.537 & -0.106 & $<2.79$ & \\
228 & SNR G323.5+00.1 & -36.514 & 0.112 & $<2.58$ & \\
229 & SNR G326.3$-$01.8 & -33.705 & -1.763 & $<2.58$ & \\
230 & SNR G327.1$-$01.1 & -32.904 & -1.106 & $<2.38$ & \\
231 & SNR G327.2$-$00.1 & -32.757 & -0.128 & $7.20\pm0.78$ (9.23) & continuum (1E 1547.0-5408); $F_{78}~\textless~1\times10^{-5}~\textrm{ph~} \textrm{cm}^{-2}~\textrm{s}^{-1}$ \\
232 & SNR G327.4+00.4 & -32.752 & 0.486 & $<2.34$ & \\
233 & SNR G327.4+01.0 & -32.631 & 1.007 & $<2.33$ & \\
234 & SNR G327.6+14.6 & -32.428 & 14.566 & $<3.00$ & SN 1006 \\
235 & SNR G328.4+00.2 & -31.589 & 0.229 & $<2.27$ & \\
236 & SNR G329.7+00.4 & -30.280 & 0.406 & $<2.16$ & \\
237 & SNR G330.0+15.0 & -30.199 & 15.534 & $<3.01$ & \\
238 & SNR G330.2+01.0 & -29.827 & 0.983 & $<2.13$ & \\
239 & SNR G332.0+00.2 & -27.954 & 0.210 & $<2.03$ & \\
240 & SNR G332.4+00.1 & -27.593 & 0.120 & $<2.00$ & \\
241 & SNR G332.4$-$00.4 & -27.572 & -0.363 &  $<2.00$ & confusion (IGR J16175-5059)\\
242 & SNR G332.5$-$05.6 & -27.421 & -5.575 & $<2.27$ & \\
243 & SNR G335.2+00.1 & -24.819 & 0.060 & $<2.00$ & confusion (IGR J16283-4838) \\
244 & SNR G336.7+00.5 & -23.245 & 0.531 & $<1.95$ & confusion (4U 1630-47) \\
245 & SNR G337.0$-$00.1 & -23.022 & -0.129 & $<1.95$ & confusion (IGR J16358-4726)\\
246 & SNR G337.2+00.1 & -22.828 & 0.055 &  $<1.95$ & confusion (4U 1630-47, IGR J16358-4726)  \\
247 & SNR G337.2$-$00.7 & -22.811 & -0.736 & $<1.86$ & \\
248 & SNR G337.3+01.0 & -22.667 & 0.962 & $<1.88$ & \\
249 & SNR G337.8$-$00.1 & -22.215 & -0.102 &  $<1.95$ & confusion (AX J163904-4642) \\
250 & SNR G338.1+00.4 & -21.900 & 0.420 & $<1.95$ & confusion (AX J163904-4642) \\
251 & SNR G338.3$-$00.0 & -21.678 & -0.079 & $<1.95$ & confusion (AX J163904-4642) \\
252 & SNR G338.5+00.1 & -21.474 & 0.067 & $<1.83$ & \\
253 & SNR G340.4+00.4 & -19.598 & 0.448 &  $<1.80$ & confusion (SNR G340.6+00.3) \\
254 & SNR G340.6+00.3 & -19.401 & 0.344 & $<1.80$ &  confusion (SNR G340.4+00.4) \\
255 & SNR G341.2+00.9 & -18.814 & 0.864 & $<1.94$ & confusion (IGR J16493-4348)\\
256 & SNR G341.9$-$00.3 & -18.140 & -0.317 &  $<1.78$ & confusion (SNR G342.0$-$00.2)\\
257 & SNR G342.0$-$00.2 & -18.057 & -0.207 & $<1.78$ & confusion (SNR G341.9$-$00.3) \\
258 & SNR G342.1+00.9 & -17.899 & 0.888 & $<1.78$ & \\
259 & SNR G343.0$-$06.0 & -17.012 & -6.035 & $<2.07$ & \\
260 & SNR G343.1$-$00.7 & -16.919 & -0.594 & $<1.92$ & confusion (IGR J17014-4306) \\
261 & SNR G343.1$-$02.3 & -16.909 & -2.310 & $<1.94$ & confusion (4U 1705-440)\\
262 & SNR G344.7$-$00.1 & -15.319 & -0.156 & $<1.71$ & \\
263 & SNR G345.7$-$00.2 & -14.273 & -0.184 & $<1.67$ & \\
264 & SNR G346.6$-$00.2 & -13.373 & -0.221 & $<1.74$ & confusion (1RXS J170849.0-400910) \\
265 & SNR G347.3$-$00.5 & -12.626 & -0.509 & $<1.69$ &  \\
266 & SNR G348.5+00.1 & -11.609 & 0.161 & $<1.69$ & \\
267 & SNR G348.5$-$00.0 & -11.403 & -0.012 & $<1.69$ & \\
268 & SNR G348.7+00.3 & -11.346 & 0.395 & $<1.52$ & \\
269 & SNR G349.2$-$00.1 & -10.870 & -0.070 & $<1.52$ & confusion (IGR J17164-3803) \\
270 & SNR G349.7+00.2 & -10.269 & 0.177 &  $<1.46$ & confusion (SNR G350.1$-$00.3)\\
271 & SNR G350.0$-$02.0 & -10.078 & -2.046 & $<1.47$ & \\
272 & SNR G350.1$-$00.3 & -10.278 & 0.248 & $<1.46$ & confusion (SNR G349.7+00.2) \\
273 & SNR G351.2+00.1 & -8.732 & 0.158 & $<1.36$ & \\
274 & SNR G351.7+00.8 & -8.296 & 0.817 & $<1.42$ & \\
275 & SNR G351.9$-$00.9 & -8.077 & -0.960 & $<1.33$ & \\
276 & SNR G352.7$-$00.1 & -7.255 & -0.120 & $<1.29$ & \\
277 & SNR G353.6$-$00.7 & -6.445 & -0.649 & $<1.27$ & \\
278 & SNR G353.9$-$02.0 & -6.059 & -2.085 & $<1.27$ & \\
279 & SNR G354.1+00.1 & -5.811 & 0.146 & $<1.32$ & confusion (GX 354-0) \\
280 & SNR G354.8$-$00.8 & -5.128 & -0.783 & $<1.22$ & \\
281 & SNR G355.4+00.7 & -4.597 & 0.727 & $<1.27$ & confusion (IGR J17315-3221) \\
282 & SNR G355.6$-$00.0 & -4.313 & -0.078 & $<1.27$ & confusion (IGR J17353-3257)\\
283 & SNR G355.9$-$02.5 & -4.054 & -2.535 & $<1.20$ & \\
284 & SNR G356.2+04.5 & -3.782 & 4.464 & $<1.29$ & \\
285 & SNR G356.3$-$00.3 & -3.701 & -0.355 & $<1.16$ & \\
286 & SNR G356.3$-$01.5 & -3.690 & -1.503 & $0.86\pm0.39$ (2.22) & \\
287 & SNR G357.7+00.3 & -2.332 & 0.348 & $<1.22$ & \\
288 & SNR G357.7$-$00.1 & -2.312 & -0.121 & $<1.12$ & \\
289 & SNR G358.0+03.8 & -2.035 & 3.803 & $<1.22$ & \\
290 & SNR G358.1+00.1 & -1.881 & 1.038 & $<1.12$ & \\
291 & SNR G358.5$-$00.9 & -1.416 & -1.002 & $<1.11$ & \\
292 & SNR G359.0$-$00.9 & -0.999 & -0.917 & $<1.11$ & confusion (SLX1744-299/300) \\
293 & SNR G359.1+00.9 & -0.903 & 0.988 & $<1.17$ & \\
294 & SNR G359.1$-$00.5 & -0.879 & -0.506 &  $<1.17$ & confusion (IGR J17446-2947) \\
\end{longtable}
\footnotetext[1]{Catalogue is sorted by source name, which is in turn,
based on Galactic coordinates.}  
\footnotetext[2]{The upper limits are given at $3\sigma$ confidence level. In case of a source detection, the significance is shown in parenthesis.}
  \footnotetext[3]{The spatial
confusion with a known hard X-ray source (shown in parenthesis) is indicated. SNRs dominated by the continuum flux are marked as well. If significance of the detection of such source is above $3\sigma$ level, an upper limit ($3\sigma$) in the 78.4 keV line is shown assuming a simple power-law continuum model.}
\end{center}

\twocolumn

\label{lastpage}

\end{document}